\documentclass[twocolumn,pr]{revtex4}

\usepackage{graphicx}
\usepackage{dcolumn}
\usepackage{bm}
\usepackage{multirow}
\usepackage{color}
\usepackage{amsmath}
\usepackage{amssymb}
\usepackage{gensymb}
\usepackage[colorlinks=true]{hyperref}

\renewcommand{\Im}{\mathop{\rm Im}}

\definecolor{mblue}{rgb}{0,0.35,0.75}

\begin{document}
\title{Exciton-phonon coupling in MoSe$_2$ monolayers}

\author{S. Shree$^1$}
\author{M. Semina$^2$}
\author{C. Robert$^1$}
\author{B. Han$^1$}
\author{T. Amand$^1$}
\author{A. Balocchi$^1$}
\author{M. Manca$^1$}
\author{E.~Courtade$^1$}
\author{X. Marie$^1$}
\author{T. Taniguchi$^3$}
\author{K. Watanabe$^3$}
\author{M. M. Glazov$^2$}
\email{glazov@coherent.ioffe.ru}
\author{B. Urbaszek$^1$}
\email{urbaszek@insa-toulouse.fr}

\affiliation{%
$^1$Universit\'e de Toulouse, INSA-CNRS-UPS, LPCNO, 135 Av. Rangueil, 31077 Toulouse, France}
\affiliation{$^2$Ioffe Institute, 194021 St.\,Petersburg, Russia}
\affiliation{$^3$National Institute for Materials Science, Tsukuba, Ibaraki 305-0044, Japan}

\begin{abstract}
We study experimentally and theoretically the exciton-phonon interaction in MoSe$_2$ monolayers encapsulated in hexagonal BN, which has an important impact on both optical absorption and emission processes. The exciton transition linewidth down to $1$~meV at low temperatures makes it possible to observe high energy tails in absorption and emission extending over several meV, not masked by inhomogeneous broadening. We develop an analytical theory of the exciton-phonon interaction accounting for the deformation potential induced by the longitudinal acoustic phonons, which plays an important role in exciton formation. The theory allows fitting absorption and emission spectra and permits estimating the deformation potential in MoSe$_2$ monolayers. We underline the reasons why exciton-phonon coupling is much stronger in two-dimensional transition metal dichalcodenides as compared to conventional quantum well structures. The importance of exciton-phonon interactions is further highlighted by the observation of a multitude of Raman features in the photoluminescence excitation experiments.
\end{abstract}

\maketitle

\section{Introduction}
Layered van der Waals materials provide a versatile platform for exploring fundamental physics and potential applications by combining different two-dimensional (2D) crystals~\cite{Novoselov:2016a,Geim:2013a,Mak:2010a, Splendiani:2010a, Wang:2012c,Mak:2016a}. Among these layered materials, transition metal dichalcogenides (TMDs) are particularly suitable for fundamental and applied optics, as they are semiconductors with a direct bandgap when thinned down to one monolayer (ML)~\cite{Splendiani:2010a,Mak:2010a}. The Coulomb interaction between conduction electrons and valence holes is strong due to 2D confinement of charge carriers, heavy effective masses and weak screening. As a result, the optical properties of monolayer TMDs  are dominated by excitons, electron-hole pairs bound by Coulomb attraction  \cite{Wang:2018a,He:2014a,Ugeda:2014a,Chernikov:2014a,Ye:2014a,Qiu:2013a,Ramasubramaniam:2012a,Wang:2015b}.\\
\indent The interaction of excitons with phonons governs many important aspects of the optical properties of 2D materials \cite{Zhang:2015a,Ribeiro:2014a,Rice:2013a,Molina:2011a,Horzum:2013a,sekine1980raman,jakubczyk2018impact} and semiconductor nano-structures in general \cite{hameau1999strong,lee1986luminescence,tatham1989time,gindele1999excitons,PhysRevB.69.035304,PhysRevB.63.155307,PhysRevLett.87.157401} including energy relaxation, dephasing and transition linewidth broadening with temperature.
Under non-resonant optical excitation high energy excitons are generated that subsequently loose energy by phonon emission, as recently discussed for MoSe$_2$ MLs~\cite{chow2017phonon}. Interactions of carriers with phonons play a key role in the exciton formation process \cite{PhysRevB.9.690,PhysRevB.50.11624,PhysRevLett.93.137401,Thilagam:2016a}. Other important signatures of the exciton-phonon interaction are single resonant and double resonant Raman scattering processes~\cite{Guo:2015a,Carvalho:2015a,Wang:2015g,Chow:2017a,molas2017raman,Soubelet:2016a}. 
Also the possible impact of polarons, electrons dressed by a phonon-cloud, on the optical properties has been discussed~\cite{Christiansen:2017a} for TMD MLs. 
There is a multitude of neutral and charged excitons in ML TMDs that give rise to complex photoluminescence spectra~\cite{Koperski:2017a}. Here phonons provide the necessary energy and in certain cases momentum for transitions between states in different valleys \cite{Dery:2015a,Dery:2016a,Lindlau:2017a}. Knowing details about the interaction of excitons with phonons is therefore important for exciton dynamics that dominates light-matter interaction in TMD MLs.\\ 
\indent Recently encapsulation in hexagonal boron nitride (hBN) of TMD monolayers has resulted in considerable narrowing of the exciton transition linewidth,  down to about 1~meV \cite{Jin:2016a,Manca:2017a,Chow:2017a,Cadiz:2017a,Ajayi:2017a,Wang:2016v}. This gives now access to interesting details of the exciton spectra. 
Here we show in photoluminescence excitation spectroscopy on monolayer MoSe$_2$ encapsulated in hBN at T$=4$~K a strong absorption tail at energies as high as 15~meV above the neutral exciton peak, well outside the transition linewidth. We demonstrate in resonant absorption experiments with a tunable laser that this high-energy tail is a very prominent feature in several samples investigated, underlining the importance of phonon-assisted exciton formation.
In temperature dependent PL experiments we show a high energy emission tail appearing as the temperature is raised from 9 to 70~K. We argue that the efficient interaction between exciton and acoustic phonons is responsible for both the broadening in absorption and the unusual evolution of the PL line with temperature observed on the same sample. Based on model calculations within an analytical approach and taking into account the deformation potential interaction between excitons and acoustic phonons, we provide a fit of our absorption and emission experiments, discuss the role of phonons in exciton formation and compare our findings to exciton-phonon interactions in quasi-2D quantum well structures. \\
\indent The paper is organized as follows: In Sec.~\ref{sec:exper} we report on experimental methods and results. Section~\ref{sec:theor} presents the model and analytical results. In Sec.~\ref{sec:disc} we compare the experiment and theory and estimate the exciton --  acoustic phonon coupling strength.  Conclusions are given in Sec.~\ref{sec:concl}.

\section{Optical Spectroscopy Results}\label{sec:exper}

\begin{figure*}
\includegraphics[width=0.80\textwidth,keepaspectratio=true]{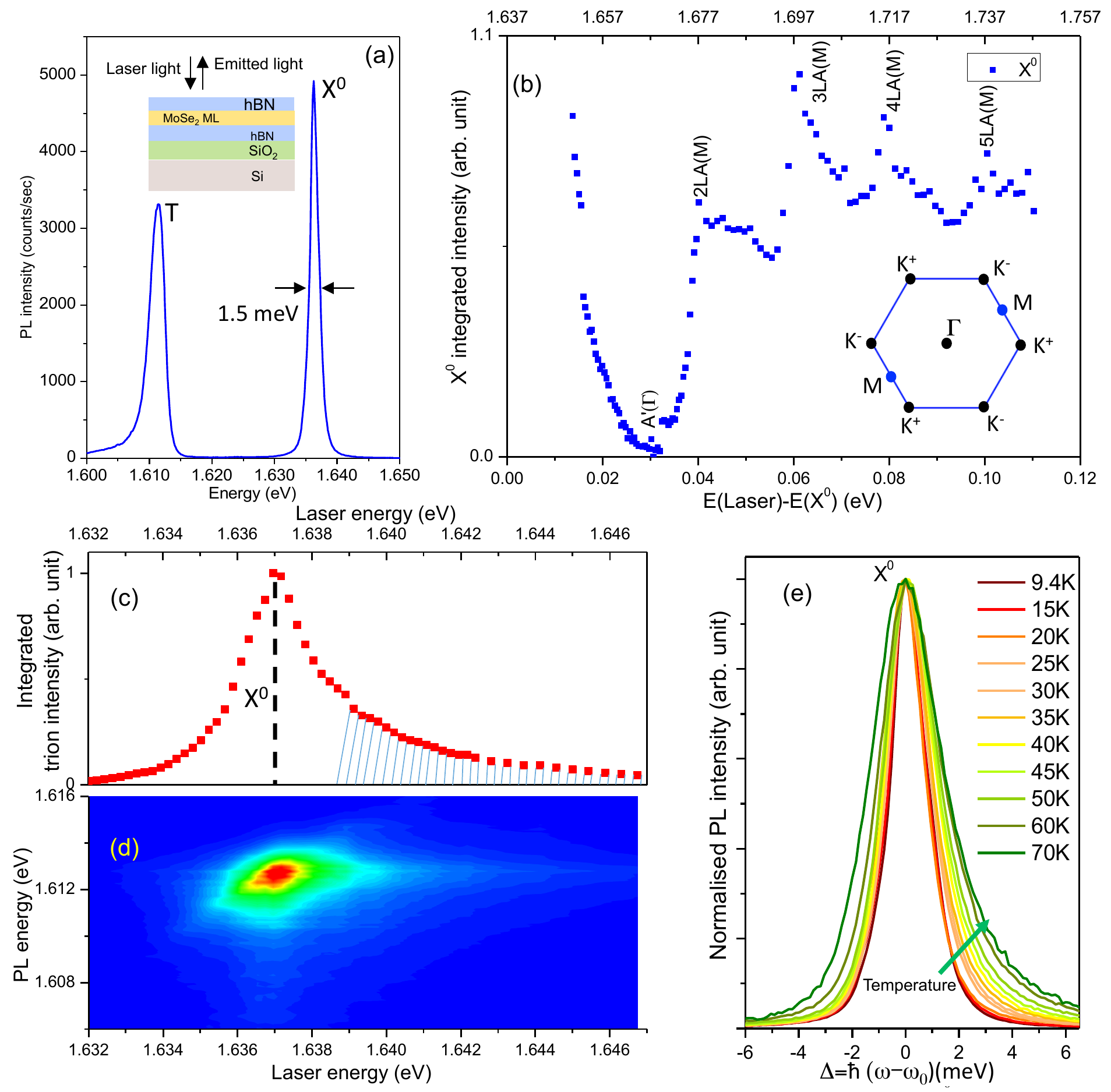}
\caption{\textbf{Optical spectroscopy results.} ML MoSe$_2$ encapsulated in hBN - \textit{sample 1} (a) Typical PL emission spectrum at T=4~K. Emission peaks at energy 1.613~eV and 1.637~eV are due to the charged (T) and the neutral exciton X$^0$. Inset of panel (a): Schematics of sample structure: MoSe$_2$ ML encapsulated in hBN, the bottom hBN layer is ($140\pm5$~nm) thick as determined by AFM, the SiO$_2$ layer is 80~nm thick. (b) PLE experiments : Integrated X$^0$ PL intensity plotted in terms of excess energy (bottom axis), defined as the difference between the laser energy (top axis) and the X$^0$ resonance. The average oscillation period corresponds to the M-point longitudinal acoustic phonon LA(M). Inset of panel (b): Schematics of 2D Brillouin zone with K, M and $\Gamma$ points (c) Integrated trion PL intensity as a function of excitation energy scanning laser across X$^0$ resonance and (d) contour plot of trion PL intensity scanning laser across X$^0$ resonance at $T=4$~K. (e) Temperature dependent PL emission at X$^0$ resonance. For better comparison of emission shape, the X$^0$ PL intensity is normalized to the maximum value and the energy shifted to compensate the temperature evolution of the bandgap. }\label{fig:fig1} 
\end{figure*}
\begin{figure*}
\includegraphics[width=0.68\textwidth,keepaspectratio=true]{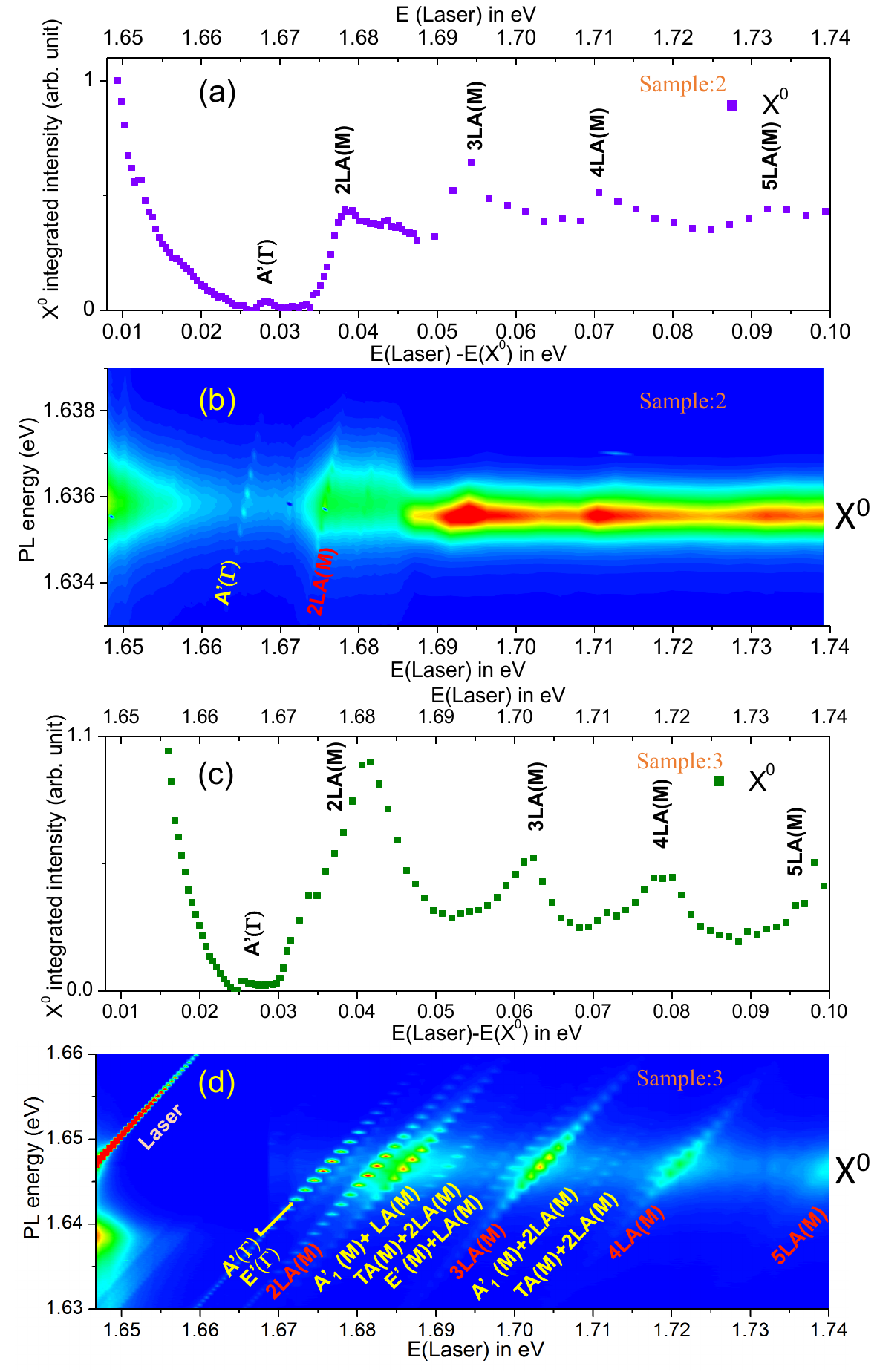}
\caption{\textit{Sample 2.} (a) PLE spectra of MoSe$_2$ ML plotted in terms of excess energy (bottom axis), defined as the difference between the laser energy (top axis) and the $X^0$ resonance. (b) Contour plot of the same data, where sharp Raman features are superimposed on broad exciton absorption features. \textit{Sample 3.} (c) Same as (a) but for different sample. (d) same as (b) but for different sample. Phonons are labelled according to \textcite{Soubelet:2016a}. In samples with more disorder and stronger localization, Raman features are more prominent as momentum conservation is relaxed \cite{Soubelet:2016a}, which leads to rich features in panel (d). In addition to PL emission, the integrated signal in panels (a) and (c) also contains contributions from Raman scattered photons if the laser energy is a multiple of a prominent phonon energy above the X$^0$ transitions. In panels (b) and (d) the Raman features are spectrally very narrow due to the narrow linewidth of the excitation laser used. }\label{fig:fig2} 
\end{figure*}
\subsection{Experimental methods}
The samples are fabricated by mechanical exfoliation of bulk MoSe$_2$ (commercially available from 2D semiconductors) and very high quality hBN crystals \cite{Taniguchi:2007a} on 80~nm-thick SiO$_2$ on a Si substrate. The experiments are carried out in a confocal microscope build in a vibration free, closed cycle cryostat with variable temperature  $T = 4-300$~K. The excitation/detection spot diameter is $\sim 1 \mu m$. 
The monolayer is excited by continuous wave (cw) Ti-Sa laser (Msquared solstis) with sub-$\mu$eV linewidth. The photoluminescence (PL) signal is dispersed in a spectrometer and detected with a Si-CCD camera. The typical laser excitation power is 3 $\mu W$.
\subsection{Experimental results}
In PL experiments shown in Fig.~\ref{fig:fig1}a  we observe well-separated ($\approx 25$ meV) emission lines for neutral (X$^0$ at $\approx1.637$~eV) and charged excitons (trions, T, at $\approx1.613$~eV) \cite{nottrion}. The transitions at low temperature in encapsulated MoSe$_2$ MLs show very narrow lines with a linewidth down to 1.2~meV \cite{Cadiz:2017a}. This gives access to physical processes previously masked by inhomogeneous contributions to the linewidth. We study now in detail quasi-resonant and resonant laser excitation of the X$^0$ transition. We start by discussing PL excitation (PLE) experiments  detecting the X$^0$ PL emission intensity as function of excitation laser energy. 
The integrated intensity of the X$^0$ transition is plotted in Fig.~\ref{fig:fig1}b. We tune the laser from 1.651~eV to 1.747~eV in 2.5~meV steps (wavelength is incremented by $\Delta \lambda=0.5$~nm). We observe oscillations of the PL intensity as a function of excitation energy, see also Fig.\ref{fig:fig2}a,c for different samples. The period of the oscillation is $\sim  20 $~meV and exactly matches the longitudinal acoustic phonon energy at the M point of the 2D hexagonal Brillouin zone, LA(M), as discussed in detail by \textcite{chow2017phonon}. As the excitation laser energy approaches the X$^0$ transition, we make an important observation: in Fig.~\ref{fig:fig1}b the PL emission intensity starts to increase considerably for excitation energies below 1.653~eV, even though the X$^0$ resonance is still $\sim$ 15 meV away and the X$^0$ PL emission line is spectrally very narrow. We verified this surprising result by performing PLE measurements on different samples of MoSe$_2$ MLs encapsulated in hBN, compared in Fig.~\ref{fig:fig2}. \\
\indent As we approach the X$^0$ transition with the laser, scattered laser light starts to obscure the PL signal on the CCD. In order to avoid this complication, we change detection scheme: We approach the X$^0$ resonance further with our laser but we detect the trion emission and assume that it is a reasonable measure of X$^0$ absorption. The excitation energy is tuned from 1.632~eV to 1.647~eV. The integrated trion emission intensity as a function of laser energy is recorded in Fig.~\ref{fig:fig1}c. This X$^0$ absorption spectrum strongly deviates from a symmetric Lorentzian lineshape, with a pronounced tail on the high-energy side of the main resonance. We have repeated this experiments for different laser powers and have reproduced each time this highly asymmetric lineshape.  \\
\indent In addition to this high energy tail in absorption, we also obtain a surprising result when studying emission at different temperature. In temperature dependent PL from 10~K to 70~K of the X$^0$, shown in Fig.~\ref{fig:fig1}d, we observe an asymmetric broadening, with stronger broadening on the high energy side, a clear deviation from a standard Lorentzian emission lineshape. \\
\indent In the following section we will explain the observed asymmetry in absorption and emission due to the efficient coupling of excitons to acoustic phonons. Our theoretical investigation is further motivated by the efficient exciton-phonon coupling observed for two additional samples in Fig.~\ref{fig:fig2}. Also for these sample we confirm the strong absorption tail at higher energy, well above the X$^0$ resonance. In Figs.~\ref{fig:fig2}a and \ref{fig:fig2}c for samples 2 and 3, respectively, this absorption tails is clearly observable up to 25~meV above the X$^0$ resonance. At higher excitation energy also for these samples we observe the multi-phonon LA(M) resonances as for sample 1 in Fig.~\ref{fig:fig1}b.

\section{Theory of exciton-phonon interaction}\label{sec:theor}

The aim of this section is to develop a simplified but analytical approach to exciton-phonon interaction in TMD monolayers. Here we argue that taking into account emission and absorption of acoustic phonons the broadening of the exciton line both in absorption at low temperature shown in Fig.~\ref{fig:fig1}c and in emission at higher temperature shown in Fig.~\ref{fig:fig1}e can be explained. We can fit the experiments to demonstrate order of magnitude agreement between our model and the measurements. For a complementary, self-consistent approach the reader is referred to \cite{Christiansen:2017a}, where a polaron framework was used.

\subsection{Deformation potential interaction}\label{subsec:dp}

In semiconductors usually two types of interaction between excitons and acoustic phonons are discussed: the deformation potential interaction and the piezoelectric interaction~\cite{gantmakher87,Cardona:2010a}. Although the latter is typically dominant for individual charge carriers~\cite{gantmakher87,Cardona:2010a,7496798}, the piezoelectric coupling is expected to be strongly suppressed for excitons in transition metal dichalcogenides due to close values of effective masses of electrons and holes (predicted by theory \cite{PhysRevB.87.115418}) and also due to thesmall exciton Bohr radius.\\
\indent The deformation potential interaction involves, in the axially symmetric approximation, longitudinal acoustic phonons. Hereafter we consider only the phonons from the MoSe$_2$ monolayer because the electron-phonon deformation potential interaction occurs in general at short range. The van der Waals interaction between the monolayer and surrounding is relatively weak and, unlike ionic or covalent bonding, is not expected to efficiently transfer deformation from one layer to another. The exciton perturbation due to the deformation potential reads
\begin{equation}
\label{Udp}
\mathcal U_{\bm q} = \sqrt{\frac{\hbar}{2\rho \Omega_{ q}S}} q b^\dag_q [e^{-\mathrm i \bm q \bm r_e} D_c - e^{-\mathrm i \bm q \bm r_h} D_v] + {\rm c.c.} \ .
\end{equation}
Here $\bm q$ is the phonon wavevector, $b^\dag_{\bm q}$ ($b_{\bm q}$) is the phonon creation (annihilation) operator, $\rho$ is the two-dimensional density of mass of the TMD ML, $\Omega_{ q} = s q$ is the phonon frequency, $s$ is the (longitudinal) sound velocity, $S$ is the normalization area, $D_{c}$ ($D_v$) are the deformation potential constants for the conduction (valence) band, ${\rm c.c.}$ stands for the complex conjugate. The matrix element describing the exciton scattering from the state with the center of mass wavevector $\bm k$ to the state with the center of mass wavevector $\bm k'$ accompanied by the phonon emission is derived from Eq.~\eqref{Udp} with the result
\begin{equation}
\label{M:em}
M^{\bm q}_{\bm k'\bm k} = \sqrt{\frac{\hbar}{2\rho \Omega_{ q}S}} q (D_c - D_v) \mathcal F(q) \delta_{\bm k, \bm k'+\bm q},
\end{equation}
where the form-factor
\begin{equation}
\label{form}
\mathcal F(q) = \int e^{-\mathrm i \bm q\bm \rho/2} \varphi^2(\rho) d\bm \rho = \frac{1}{\left[1+ \left(\frac{q a_B}{4}\right)^2 \right]^{3/2}},
\end{equation}
is introduced. Here and in what follows we assume equal electron and hole effective masses, $\varphi(\rho) = \sqrt{2/(\pi a_B^2)}\exp{(-\rho/a_B)}$ is the ground state envelope function (taken in the hydrogenic form), $a_B$ is the effective exciton Bohr radius. For relatively small phonon wavevectors $q a_B \ll 1$ the form-factor is not sensitive to the shape of the envelope function.

\subsection{Exciton-phonon scattering}\label{subsec:scatt}
\begin{figure*}
\includegraphics[width=0.85\textwidth,keepaspectratio=true]{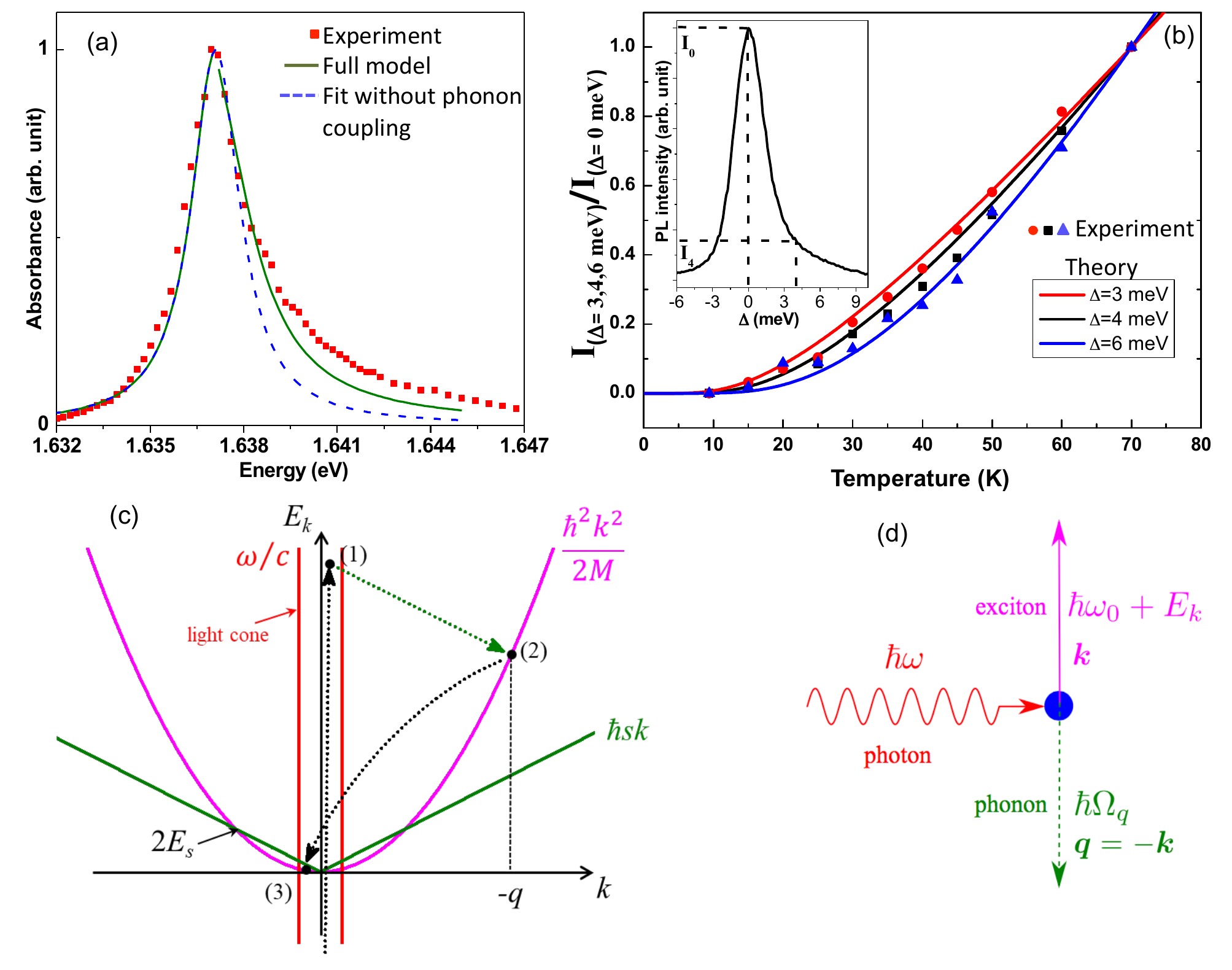}
\caption{ (a) Absorbance at the A:1s exciton in MoSe$_2$ monolayer. Experimental data same as Fig.~\ref{fig:fig1}c - squares, full theory by solid green curve, Eqs.~\eqref{Abs:tot} and \eqref{gamma:ph:small:T}, blue dashed curve shows the result in the absence of phonon-assisted transitions. $\Gamma_0=\Gamma=0.47$~meV, $|D_c-D_v|=11$~eV, other parameters are from the Tab.~\ref{tab:parameters}.  in the calculation $T=4$~K. (b) Normalized to the value at $T=70$~K with the $T\to0$ offset subtracted intensity of PL emission measured (symbols) and calculated after Eq.~\eqref{I:PL:simpl1} (solid lines) as a function of temperature at a fixed detunings of $\Delta=3$ (red), $4$ (black), and $6$~meV (blue). The parameters of the calculation are the same as in (a). (c) Schematics illustrating exciton and phonon dispersion, the intersection gives the energy scale $2Ms^2$, for $E_k\gg Ms^2$ the exciton-phonon interaction is quasi-elastic. Steps (1), (2), and (3) describe, respectively, photon absorption with exciton formation in a virtual state, exciton-phonon scattering resulting in a real final state for exciton (on the exciton dispersion, 2), and relaxation toward the radiative states inside the light cone (3). (d) Sketch of the photon absorption accompanied by the phonon emission. Exciton and phonon wavevectors have same absolute value but opposite directions. }\label{fig:fig3} 
\end{figure*}
In this subsection we study the phonon-induced broadening of exciton resonance in the simplest possible approach based on the Fermi golden rule. To that end we introduce the out-scattering rate from the exciton state with the  wavevector $\bm k$ and kinetic energy $E_k = \hbar^2 k^2/2M$, where $M$ is the translational mass of the exciton, due to emission of the acoustic phonons:
\begin{equation}
\label{lifetime}
\frac{1}{\tau_{k}} = \frac{2\pi}{\hbar} \sum_{\bm k'\bm q} |M^{\bm q}_{\bm k'\bm k}|^2 \delta (E_k - E_{k'} - \hbar\Omega_q).
\end{equation}
For $E_k, E_{k'}\gg E_s$, where $E_s = Ms^2$, the exciton-phonon scattering is quasi-elastic~\cite{gantmakher87} (this point is discussed in detail below in Sec.~\ref{subsec:pert}), thus we can omit the phonon energy, $\hbar\Omega_q$, in the energy conservation $\delta$-function. Assuming that typical phonon and exciton wavevectors are smaller than the inverse Bohr radius, $k,k' \ll a_B^{-1}$ (this condition is fulfilled at all realistic temperatures), we put the form-factor $\mathcal F\equiv 1$ and obtain from Eq.~\eqref{lifetime}:
\begin{equation}
\label{lifetime:2}
\frac{1}{\tau_{k}} = 4 \frac{\sqrt{2E_s^3E_k}}{\hbar \mathcal E}.
\end{equation}
Here
\begin{equation}
\label{not}
\mathcal E^{-1} = \frac{(D_c-D_v)^2}{4\pi \hbar^2 \rho s^4}.
\end{equation}

Since exciton-phonon scattering is quasi-elastic both absorption and emission processes are possible because the phonon energy $\hbar \Omega_q$ is typically much smaller than $k_B T$ with $T$ being the temperature and $k_B$ the Boltzmann constant. Thus, for a thermalized exciton gas one has 
\begin{equation}
\label{lifetime:Temp}
\frac{1}{\tau_{T}} = \frac{2\pi}{\hbar} \sum_{\bm k'\bm q} |M^{\bm q}_{\bm k'\bm k}|^2 (1+2n_q) \delta ({E_k - E_{k'}}),
\end{equation}
where 
\begin{equation}
\label{bose}
n_q = \frac{1}{\exp{\left( \frac{\hbar\Omega_q}{k_B T}\right)} - 1}
\end{equation}
is the phonon occupation numbers. For $\hbar\Omega_q \ll E_k,E_{k'} \sim k_B T$ we have $n_q = k_B T/\hbar\Omega_q \gg 1$ and, as it follows from Eq.~\eqref{lifetime:Temp}~\footnote{In Ref.~\cite{PhysRevB.85.115317} analogous expression, Eq. (25), was derived for the electron-phonon scattering.}
\begin{equation}
\label{lifetime:Temp:1}
\frac{1}{\tau_{T}} = \frac{4\pi}{\hbar} \frac{E_s k_B T}{\mathcal E}.
\end{equation}
The phonon-induced exciton linewidth can be roughly estimated making use of Eqs.~\eqref{lifetime:Temp}, \eqref{lifetime:Temp:1} with the result
\begin{equation}
\label{broadening}
\frac{\hbar}{\tau_T} =c_1 T, \quad  c_1 = 4\pi \frac{E_s k_B T}{\mathcal E}.
\end{equation}
The parameter $c_1$ can be accessed in experiments~\cite{Selig:2016aa,jakubczyk2016radiatively,Cadiz:2017a}, see also this work, Appendix~\ref{appendix:c1:exp}. Since the material parameters and the speed of sound are known to a large extent, the experimentally determined value of $c_1^{exper}$ allows us to estimate the strength of the exciton-phonon interaction. Relevant parameters of exciton-phonon interaction are summarized in Tab.~\ref{tab:parameters}. The analysis in Appendix~\ref{appendix:c1:exp} demonstrates that the accuracy of determination of $c_1$ parameter is not extremely high because at moderate to high temperatures the broadening can be  dominated by the optical phonons and acoustic phonons at the Brillouin zone edge. Therefore the values $c_1^{exper}$ and the deduced deformation potentials can be considered as order-of-magnitudes estimates only.

\begin{table}[t] 
\caption{Material parameters}
\label{tab:parameters}
\begin{tabular}{c|c|c|c|c}
\hline
Property & MoSe$_2$ & MoS$_2$ & WSe$_2$ & WS$_2$\\
\hline
$s$ ($10^5\times$cm/s)\footnote{From Ref.~\cite{PhysRevB.90.045422}.} & $4.1$ & $6.6$ & $3.3$ & $4.3$\\
$\rho$ ($10^{-7}\times$g/cm$^2$)\footnote{Calculated from the data on bulk crystals as $\rho=\rho_{bulk}c/2$, where $c$ is the lattice constant for the 2H polytype.} & $4.46$ & $3.11$\footnote{From Ref.~\cite{PhysRevB.85.115317}.} & $6.04$ & $4.32$\\
$c_1^{exper}$ ($\mu$eV/K) & $91$\footnote{From Ref.~\cite{Selig:2016aa}}~,~${52\ldots 88}$\footnote{This work} & $45$~\footnote{From Ref.~\cite{PhysRevLett.116.127402}}~,~$70$~\footnote{From Ref.~\cite{Cadiz:2017a}} & ${60}$~\footnote{From Ref.~\cite{Moody:2015a}.} & $28^{\rm d}$\\
$|D_c-D_v|^{fit}$ (eV)\footnote{Fit to experimental data on $c_1$ (thermal broadening of exciton resonance) after Eq.~\eqref{broadening}.} & ${5\ldots 6.5}$ & $5.5^{\rm f}$~,~$7.7^{\rm g}$, $10.5$\footnote{From experimental data in Ref.~\cite{2053-1583-2-1-015006} on strain tuning of optical resonances.} & $5.4$\footnote{From experimental data in Ref.~\cite{2053-1583-3-2-021011} on strain tuning of optical resonances.} & $3.5$\\
\hline
\end{tabular}
\end{table}

\subsection{Phonon-assisted absorption and high-energy tail. Perturbative derivation}\label{subsec:pert}

In this paragraph we aim to discuss exciton formation, assisted by phonons, for laser excitation energies below the first LA(M) resonance i.e. closer than 19~meV to the neutral exciton resonance.
In absence of exciton-phonon interaction the absorbance of the transition metal dichalcogenide monolayer at the normal incidence of radiation in the spectral vicinity of exciton resonance is described by the Lorentzian profile
 \begin{equation}
\label{abs:dir:1}
\mathcal A_{0}(\omega) =  \frac{2\Gamma_0\Gamma}{(\hbar\omega - \hbar\omega_0)^2 + (\Gamma+\Gamma_0)^2}. 
\end{equation}
Here $\omega$ is the incident light frequency, $\omega_0$ is the exciton resonance frequency, $\Gamma_0$ is the exciton radiative decay rate, $\Gamma$ is the non-radiative decay rate of the exciton. This expression can be derived within the non-local dielectric response theory~\cite{Glazov:2014a,Ivchenko:2005a} by calculating the amplitude reflection $r(\omega)$ and transmission $t(\omega) = 1+r(\omega)$ coefficients of the monolayer and evaluating the absorbance as $\mathcal A_0(\omega) = 1- |r(\omega)|^2 - |t(\omega)|^2$. Hereafter we disregard the effects related with the light propagation in the van der Waals heterostructure~\cite{PhysRevMaterials.2.011001}. The estimates performed for our structures show that the asymmetry in the shape of the $\mathcal A_0(\omega)$ related to the light propagation effects is negligible.

To simplify derivations we consider in a first approach in what follows the limit of $\Gamma_0\ll \Gamma$, which allows us to take into account the light-matter interaction perturbatively. In this limit we can arrive at Eq.~\eqref{abs:dir:1} making use of the Fermi's golden rule and evaluating the rate of the transitions to the excitonic state with the in-plane wavevector $\bm k=0$ as
\begin{equation}
\label{abs:dir}
\mathcal A_{0} = \frac{2\pi}{\hbar} |M_{opt}|^2 \delta (\hbar\omega - \hbar\omega_0).\end{equation}
Here $|M_{opt}|^2$ is the squared matrix element of the optical transition to the $\bm k=0$ state normalized per flux of the incident photons. The $\delta$-function ensuring the energy conservation is broadened by non-radiative processes
\[
\delta(\hbar\omega - \hbar\omega_0) = \frac{1}{\pi} \frac{\Gamma}{(\hbar\omega - \hbar\omega_0)^2 + \Gamma^2}. 
\]
The radiative broadening of the exciton resonance can be introduced as $\Gamma_0= |M_{opt}|^2/\hbar$.
Within second-order perturbation theory the phonon-assisted absorption [schematically shown in Fig.~\ref{fig:fig3}, panels (c) and (d)] is possible for $\hbar\omega > \hbar\omega_0$ via a two step process: First the exciton is created in the intermediate (virtual) state with $\bm k \simeq 0$ and the emission of the phonon takes place as a second step. Correspondingly, the contribution to absorbance due to this two-step process reads
\begin{equation}
\label{abs:indir}
\mathcal A_{1} = \frac{2\pi}{\hbar} \sum_{\bm k, \bm q} \left|\frac{M^{\bm q}_{\bm k, \bm 0}M_{opt}}{\Delta} \right|^2 \delta(\Delta - E_{\bm k} - \hbar\Omega_{\bm q}).
\end{equation}
Here 
\begin{equation}
\label{not:1}
\Delta = \hbar\omega - \hbar\omega_0, \end{equation}
is the detuning between the photon energy and the exciton resonance energy.

It follows from the momentum conservation law that the phonon wavevector $\bm q = - \bm k$ as the total linear momentum of the electron-hole pair is equal to the in-plane photon momentum, that is close to zero at the normal incidence of radiation. The energy conservation law therefore reads
\begin{equation}
\label{cons}
\frac{\hbar^2 k^2}{2M} + \hbar s k = \Delta.
\end{equation}
Equation~\eqref{cons} can be readily solved to obtain the phonon energy in the form ($q=k$)
\begin{equation}
\label{solution}
\hbar\Omega_k = E_s \left( \sqrt{1+ \frac{2\Delta}{E_s}} - 1\right) \approx
\begin{cases}
\Delta, \quad \Delta \ll E_s,\\
\sqrt{2\Delta E_s}, \quad \Delta \gg E_s.
\end{cases}
\end{equation}
For typical parameters $E_s \sim 100$~$\mu$eV$\ll \Delta$ and all energy is transferred to the recoil exciton, leaving the phonon with a small energy $\sqrt{2\Delta E_s} \ll \Delta$. An illustration of the exciton and phonon dispersions is shown in Fig.~\ref{fig:fig3}c. Summation over $\bm k$ in Eq.~\eqref{abs:indir} can be performed for any ratio of $\Delta$ to $E_s$ with the result
\begin{equation}
\label{abs:indir:1}
\mathcal A_{1} = 2\pi \Gamma_0 \frac{\mathcal F^2(\Omega_k/s)}{\mathcal E} \mathcal P\left(\frac{\Delta}{E_s}\right) \Theta(\Delta),
\end{equation}
where $\Theta(\Delta)$ is the Heaviside $\Theta$-function and the function $\mathcal P(x)$ is defined as
\begin{equation}
\mathcal P(x) = \frac{1}{\sqrt{1+2x}} \left(\frac{\sqrt{1+2x} - 1}{x} \right)^2 \approx
\begin{cases}
1, \quad x\ll 1,\\
\sqrt{2/x^3}, \quad x\gg 1.
\end{cases}
\end{equation}
For the experimentally relevant situation where $x=\Delta/E_s \gg 1$ the function $\mathcal P(x) \approx\sqrt{2}/x^{3/2}$ and $\mathcal A_{1}$ in Eq.~\eqref{abs:indir:1} reduces to 
\begin{equation}
\label{abs:indir:2}
\mathcal A_{2} = 2\sqrt{2}\pi \Gamma_0 \frac{\mathcal F^2(\Omega_k/s)}{\mathcal E} \left(\frac{E_s}{\Delta}\right)^{3/2}\Theta(\Delta).
\end{equation}
Expression~\eqref{abs:indir:2} describes the tail of absorption at $\hbar\omega-\hbar\omega_0 \equiv \Delta  \gg \Gamma,\Gamma_0,E_s$, $\Omega_k$ is given by Eq.~\eqref{solution}, similarly to predicted and observed in Ref.~\cite{Christiansen:2017a}. Accounting for both the phonon emission and absorption processes gives rise to an additional factor of $(1+2n_k)$ in Eq.~\eqref{abs:indir:2}.

For the purpose of direct comparison with experiment, it is convenient to obtain a model expression for the absorption spectrum which accounts for both the (no-phonon) direct and indirect (phonon-assisted) processes. To that end we employ the self-consistent Born approximation~\cite{JPSJ.36.959} for the exciton-phonon interaction, see Appendix~\ref{appendix:scba} for details, and arrive at
\begin{equation}
\label{Abs:tot}
\mathcal A (\hbar\omega) = \frac{2\Gamma_0[\Gamma+\gamma_{ph}(\Delta)]}{\Delta^2 + [\Gamma+\Gamma_0+\gamma_{ph}(\Delta)]^2},
\end{equation}
where the acoustic phonon scattering rate is: 
 \begin{multline}
\label{gamma:ph:small:T}
\gamma_{ph}(\Delta) = \\
\coth{\left( \frac{\hbar\Omega_k}{2k_B T}\right)}\frac{\pi \mathcal F^2(\Omega_k/s) \Delta^2  }{\mathcal E} \mathcal P\left(\frac{\Delta}{E_s}\right) \Theta(\Delta).
\end{multline}
In Eqs.~\eqref{Abs:tot} and \eqref{gamma:ph:small:T} both phonon absorption and emission processes are included. Strictly speaking, the self-consistent Born approximation is not applicable for $\Delta \to 0$, because the smoothing of the Heaviside function should be also taken into account. The fully self-consistent approach such as developed in Ref.~\cite{Christiansen:2017a} should then be applied. Here, the condition $\Gamma_0\ll \Gamma$ is relaxed that is why $\Gamma_0$ is present also in the denominator of Eq.~\eqref{Abs:tot}. The results of calculation of $\mathcal A(\hbar\omega)$ are shown in Fig.~\ref{fig:fig3}a and fit very well the measured absorption spectra, they are discussed further in Sec.~\ref{sec:disc}.

\subsection{Temperature-dependent photoluminescence}\label{subsec:PL}

In the photoluminescence process the exciton formed non-resonantly relaxes in energy towards small $k$-states.  For a monolayer in free space the light cone is defined as $q_{phot} = \omega/c$, $c$ is the speed of light, and when $k(\omega)\leqslant q_{phot}$ the exciton can recombine by emitting a photon. In the absence of exciton-phonon interaction the photoluminescence spectrum has a simple Lorentzian form, cf. Eq.~\eqref{abs:dir:1}. The exciton-phonon interaction makes states outside the light cone optically active as well. For example, an exciton with the wavevector $\bm k$ can decay by emitting a phonon with wavevector $\bm q = \bm k$ and in addition a photon along the monolayer normal i.e. $q_{phot}=0$. For experiments at finite temperature not only the phonon emission, but also phonon absorption processes become important. \\
\indent In general calculation of the photoluminescence spectrum requires detailed analysis of the exciton energy relaxation and recombination processes~\cite{deych:075350,PhysRevB.53.15834}. Here, to simplify the consideration as we aim for a qualitative comparison with our experimental data, we assume that the excitons are thermalized and their occupation numbers are described by the Boltzmann distribution
\begin{equation}
\label{fk:exc}
f_k = \exp{\left(\frac{\mu- E_k}{k_B T}\right)},
\end{equation}
where $\mu<0$ is the chemical potential of the exciton gas. The increase in temperature mainly results in occupation of higher energy excitonic states, which due to the quasi-elastic character of the exciton-phonon interaction gives rise to the high-energy wing in the photoluminescence which is so striking for the experiments shown in Fig.~\ref{fig:fig1}e. Denoting, as in Sec.~\ref{subsec:pert}, by $\hbar\omega$ the energy of the emitted photon, by $\hbar\omega_0$ the energy of the exciton at $k=0$, and by $\Delta$ the detuning, Eq.~\eqref{not:1}, we arrive at the photon emission rate in the form [cf.~\eqref{abs:indir}]
\begin{equation}
\label{I:PL:simpl}
I(\omega) \propto \sum_{\bm k, \bm q} f_k (1+2n_q) \left|\frac{M^{\bm q}_{\bm 0, \bm k}M_{opt}}{\Delta} \right|^2  \delta(\Delta - E_{\bm k}).
\end{equation}
Equation~\eqref{I:PL:simpl} holds for $\Delta \gg \Gamma$, in the energy conservation $\delta$-function the phonon energy is omitted. The calculation shows that the photoluminescence spectrum at the positive detunings related to the phonon-assisted emission takes the form
\begin{equation}
\label{I:PL:simpl1}
I(\omega) \propto e^{-\Delta/k_BT} \frac{k_B T E_s}{\Delta^2} \mathcal F^2\left(\frac{\sqrt{2\Delta E_s}}{\hbar s}\right), \quad \Delta \gg \Gamma.
\end{equation}
The coefficient in Eq.~\eqref{I:PL:simpl1} is temperature and detuning independent. For given detuning $\Delta>0$ the relative photoluminescence intensity drastically increases with temperature mainly due to the presence of excitons with higher energies. 

At the same time, the photoluminescence spectrum at $\omega<\omega_0$ (negative detuning $\Delta<0$) is practically unaffected by the temperature increase. Indeed, to emit a photon with energy significantly (by more that $\hbar \Gamma$) lower than $\hbar\omega_0$ a phonon with the energy $\hbar\Omega_q \geqslant |\Delta|$ should be emitted as well. The rate of this process is given by
\begin{equation}
\label{I:PL:simpl:neg}
I(\omega) \propto \sum_{\bm k, \bm q} f_k (1+n_q) \left|\frac{M^{\bm q}_{\bm 0, \bm k}M_{opt}}{\Delta} \right|^2  \delta(\Delta - E_{\bm k}+\hbar\Omega_q).
\end{equation}
Due to the energy conservation law the process is allowed for small absolute values of detunings only $|\Delta|\leqslant E_s/2$, see Fig.~\ref{fig:fig3}. Thus, for the experimentally relevant parameters the photoluminescence increase at negative detunings is not expected in the theory, in agreement with the observation of an asymmetric lineshape in Fig.~\ref{fig:fig1}e. Note, that this is somewhat different from the optical spectra in quantum dot structures where the exciton energy spectrum is discrete and exciton-phonon coupling is essentially inelastic. Thus, in quantum dots deviations from Lorentzian line-shapes with wings at low and high energy can appear \cite{PhysRevB.63.155307,PhysRevB.69.035304}.
 
\section{Discussion}\label{sec:disc}

Figure~\ref{fig:fig3}a (same data as Fig.~\ref{fig:fig1}c) shows the X$^0$ absorption measured via the integrated trion PL intensity with a very prominent high energy tail. The results of our model calculations using Eqs.~\eqref{Abs:tot} and \eqref{gamma:ph:small:T} for the best fit parameters (given in the caption) are shown by the red solid line. For comparison, blue dashed curve shows the symmetric Lorentzian which describes the low-energy wing of the absorption ($\hbar\omega < \hbar\omega_0$), but does not reproduce the high energy wing. In contrast, the full calculation fits the spectrum and reproduces the observed absorption asymmetry very well, giving a strong indication that including exciton-phonon coupling is important for describing the absorption process.\\
\indent The free parameters in our fit are the radiative, $\Gamma_0$, and nonradiative, $\Gamma$ damping of the exciton as well as the difference of the deformation potentials for the electron and hole, $|D_c - D_v|$. The values used for $\Gamma_0=\Gamma\simeq 0.5$~meV is in order of magnitude agreement with results from time resolved spectroscopy reporting hundreds of fs to few ps lifetimes \cite{pollmann:2015a,jakubczyk2016radiatively,Robert:2016a}. The best fit value of $|D_c - D_v|$ of $11$~eV is somewhat larger than the literature and experimental data summarized in Tab.~\ref{tab:parameters}. The discrepancy can originate from the simplification of the model, particularly, the absence of the coupling with optical phonons as well as with phonons at the Brillouin zone edge, as well as from certain inaccuracy in determination of the deformation potentials, see Appendix~\ref{appendix:c1:exp} for a brief discussion of accuracy of fit of the temperature-induced broadening. Additionally, the presence of free charge carriers may also provide a contribution to the high-energy wing in absorption due to the possibility to relax the momentum conservation law. We also stress that the absorption (Figs.~\ref{fig:fig1}c and \ref{fig:fig3}a) is detected via the photoluminescence of the trion, therefore, efficiency of the relaxation pathway may play a role as well.\\
\indent Similarly to the high energy tail in the absorption, the calculations predict the same asymmetry in the photoluminescence emission spectrum. As described in Sec.~\ref{subsec:PL} this is because the excitons with wavevectors $k\gtrsim q_{phot}=\omega/c$ become optically active due to the phonon-assisted processes. The increase in the temperature gives rise to the increase of the occupancy of excitonic states at larger wavevectors. This is exactly what is observed in Fig.~\ref{fig:fig1}(e). Figure~\ref{fig:fig3}b shows the intensity of photoluminescence at three values of the positive detuning $\Delta=3$, $4$, and $6$~meV (all on the high energy wing) as a function of temperature. The experimental data are in reasonable agreement with the calculation after Eq.~\eqref{I:PL:simpl1} and indeed confirm the importance of the phonons in the photoluminescence emission. \\
\indent Finally, we stress that the exciton interaction with acoustic phonons is particularly strong in TMD MLsand, generally, in two-dimensional crystals as compared with conventional quantum well structures such as GaAs/AlGaAs. This is because the density of states of two-dimensional phonons $\propto q$ is greatly enhanced compared to that of the bulk phonons $\propto q^2$. This results in about an order of magnitude enhancement (see Appendix~\ref{appendix:comp} for details and further discussion) and accounts for the observation of the high-energy tails in exciton absorption in TMD monolayers.

\section{Conclusion}\label{sec:concl}
Our combined experimental and theoretical study sheds light on the strong impact of the exciton-acoustic phonon interaction in TMD monolayers on exciton formation and recombination. We interpret strong absorption above the exciton resonance in terms of phonon-assisted exciton formation. Asymmetric lineshapes with high-energy tails in emission are observed in temperature dependent experiments. Also this unusual emission lineshape is due to deformation potential exciton-acoustic phonon coupling, which is enhanced in TMD monolayers as compared to conventional quantum well structures, and can be fitted by our simplified analytical theory.\\

\indent \emph{Acknowledgements.---}  
We are grateful to A. Knorr for discussions. We acknowledge funding from ANR 2D-vdW-Spin and ANR VallEx, ITN Spin-NANO Marie Sklodowska-Curie grant agreement No 676108 and ITN 4PHOTON Nr. 721394, and Labex NEXT project VWspin. X.M. also acknowledges the Institut Universitaire de France. Growth of hexagonal boron nitride crystals was supported by the
Elemental Strategy Initiative conducted by the MEXT, Japan and the CREST
(JPMJCR15F3), JST. M.A.S. and M.M.G. acknowledge partial support from LIA ILNACS, RFBR projects 17-02-00383, 17-52-16020 and RF President Grant MD-1555.2017.2. 

\appendix

\section{Thermal broadening of exciton line}\label{appendix:c1:exp}

\begin{figure}[t]
\includegraphics[width=\linewidth]{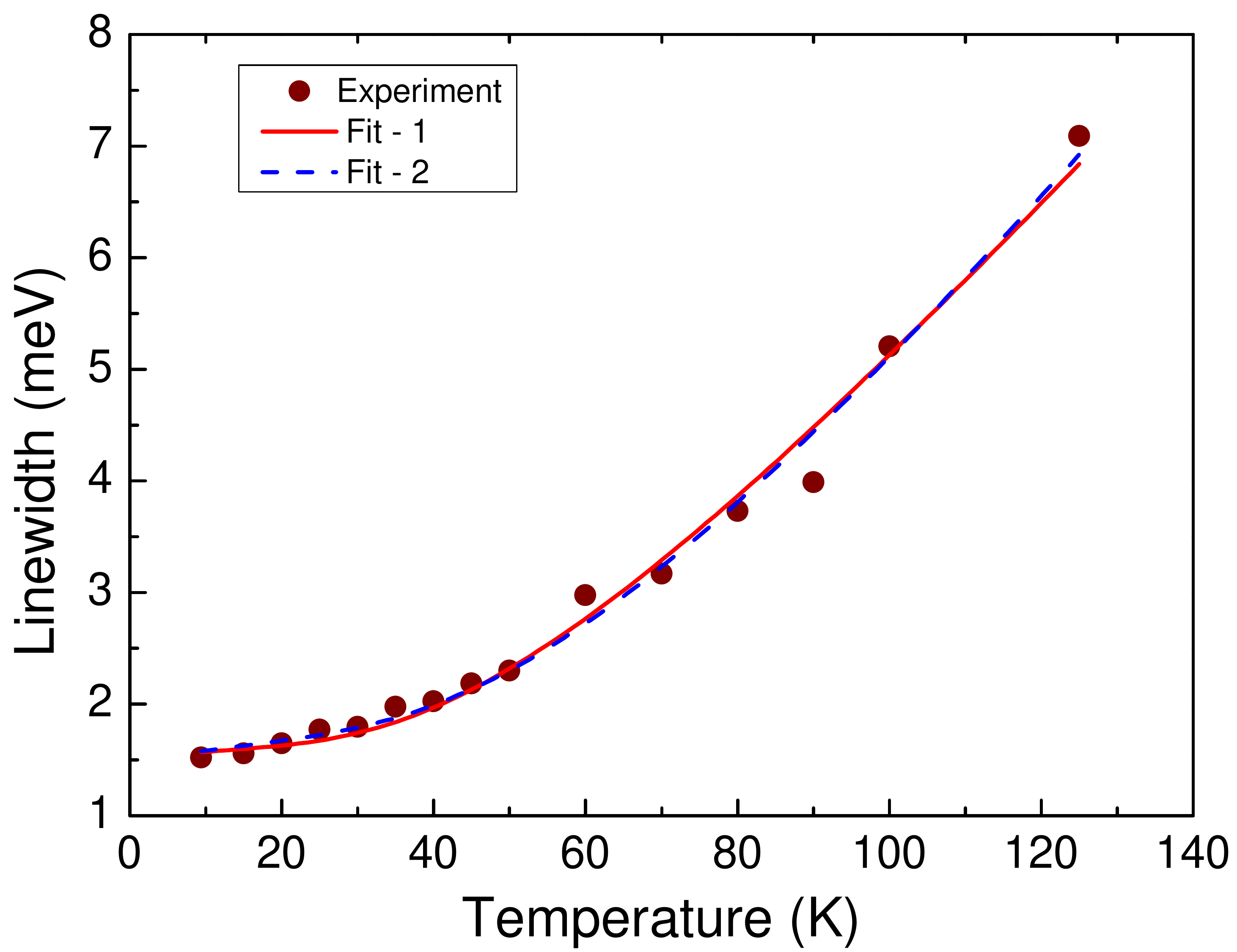}
\caption{Exciton linewidth (2$\times$ half width at half maximum at \textit{low} energy side) as a function of temperature extracted from the PL data in Fig.~\ref{fig:fig1}e (points). Fit - 1 (red solid curve) is performed by Eq.~\eqref{fit:T} with the parameter set $\gamma(0) = 1.52$~meV, $c_1=52$~$\mu$eV/K, $c_2=11.52$~meV and $\hbar\Omega_0=13.4$~meV.  Fit - 2 (blue dashed curve) is performed by Eq.~\eqref{fit:T} with the parameter set $\gamma(0) = 1.49$~meV, $c_1=88$~$\mu$eV/K, $c_2=14.9$~meV and $\hbar\Omega_0=15.5$~meV.}\label{fig:linewidth}
\end{figure}

Figure~\ref{fig:linewidth} presents the dependence of the exciton linewidth in photoluminescence (measured as full width at half maximum) on the sample temperature. Two fits are performed with very close values of parameters according to the general expression~\cite{PhysRevLett.115.267402,Selig:2016aa,PhysRevLett.116.127402}
\begin{equation}
\label{fit:T}
\gamma(T) \equiv \frac{\hbar}{\tau_T} = \gamma(0) + c_1 T + \frac{c_2}{\exp{\left(\frac{\hbar\Omega_0}{k_B T}\right)}-1}.
\end{equation}
Here $\gamma(0)$ is the zero-temperature linewidth not related to the exciton-phonon interaction, the $T$-linear term is due to the exciton-acoustic phonon scattering, Eq.~\eqref{broadening}, and the last term describes the interaction of excitons with optical phonons or acoustic phonons at the Brillouin zone edges, $\hbar\Omega_0$ is the effective phonon energy. Reasonable fit quality can be achieved for the values of $c_1$ which differ by a factor of $2$, compare Fit - 1 and Fit - 2 parameter sets in the figure caption.

\section{Nonperturbative derivation}\label{appendix:scba}

For comparison with experiment it is instructive to derive the expression for the indirect absorption spectrum relaxing the condition $\Delta \gg \Gamma$. To that end we calculate the self-energy of the exciton with $\bm k=0$ with account for the exciton-phonon interaction (allowing for, as before, the processes with the phonon emission only). In the self-consistent Born approximation it takes the form [cf. Ref.~\cite{Christiansen:2017a}]
\begin{equation}
\label{self}
\Sigma_{\bm 0}(\hbar \omega)  = \sum_{\bm k, \bm q} |M_{\bm k,\bm 0}^{\bm q}|^2 \mathcal G_{\bm k} (\hbar\omega - \hbar\Omega_q),
\end{equation}
where 
\begin{equation}
\label{ret}
\mathcal G_{\bm k} (\hbar\omega) = \frac{1}{\hbar\omega - \hbar\omega_0 - E_k - \Sigma_{\bm k}(\hbar\omega)}
\end{equation} is the ``dressed'' and
\begin{equation}
\label{bare}
G_{\bm k} (\hbar\omega) = \frac{1}{\hbar\omega - \hbar\omega_0 - E_k + \mathrm i \Gamma},
\end{equation}
is the bare retarded Greens function of the exciton in the state $\bm k$. In derivation of Eq.~\eqref{self}, \eqref{ret} we assumed that the photon damping is negligible and disregarded the diagrams with the self-crossings~\footnote{This is correct for not too small $\Delta$ where the final exciton state with the energy $E_k$ is well-defined and the standard criteria of applicability of kinetic equation is fullfilled.}. Furthermore, we assume that the phonon-induced renormalization of the (real) part of the exciton frequency are already included in $\omega_0$ and $E_k$ so that $\Sigma_{\bm k}(\hbar\omega)$ is purely imaginary. For $|\Im \Sigma_{\bm k}(\hbar\omega)|\ll E_k$ we obtain from Eqs.~\eqref{self} and \eqref{ret}
\begin{multline}
\label{self:1}
-\Im \Sigma_{\bm 0}(\hbar \omega) = \Gamma +  \pi \sum_{\bm k, \bm q} |M_{\bm k,\bm 0}^{\bm q}|^2 \delta(\hbar\omega - \hbar\omega_0 - \hbar\Omega_q)\\
= \Gamma + \gamma_{ph}(\Delta).
\end{multline}
The phonon-induced contribution can be calculated similarly to derivation of Eq.~\eqref{abs:indir:1} with the result
\begin{equation}
\label{gamma:ph:small}
\gamma_{ph}(\Delta) =  \frac{\pi \mathcal F^2(\Omega_k/s) \Delta^2  }{\mathcal E} \mathcal P\left(\frac{\Delta}{E_s}\right) \Theta(\Delta).
\end{equation}
With account for finite temperature effects (both emission and absorption) the contribution $\gamma_{ph}(\Delta)$ can be recast as
 \begin{multline}
\label{gamma:ph:small:T:App}
\gamma_{ph}(\Delta) =  (1+2n_q) \frac{\pi \mathcal F^2(\Omega_k/s) \Delta^2  }{\mathcal E} \mathcal P\left(\frac{\Delta}{E_s}\right) \Theta(\Delta) \\
= \coth{\left( \frac{\hbar\Omega_k}{2k_B T}\right)}\frac{\pi \mathcal F^2(\Omega_k/s) \Delta^2  }{\mathcal E} \mathcal P\left(\frac{\Delta}{E_s}\right) \Theta(\Delta),
\end{multline}
with $\hbar\Omega_k$ given by Eq.~\eqref{solution}.

\begin{figure}[t]
\includegraphics[width=\linewidth]{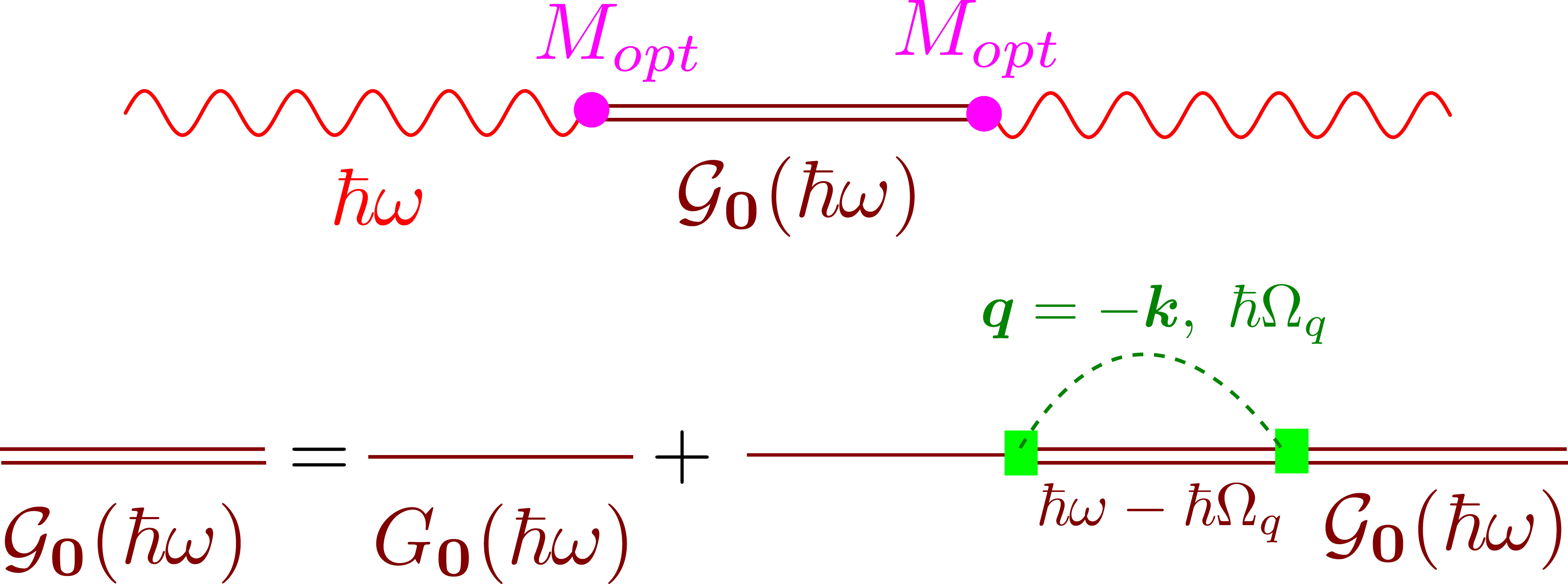}
\caption{Diagrams describing the photon absorption (top) and exciton dressed Greens function (bottom). }\label{fig:diag}
\end{figure}

The absorbance by the exciton is given by~
\begin{equation}
\label{Abs:tot:App}
\mathcal A (\hbar\omega) = -  2\Gamma_0 \Im \mathcal G_{\bm 0}(\hbar\omega) = \frac{2\Gamma_0[\Gamma+\gamma_{ph}(\Delta)]}{\Delta^2 + [\Gamma+\gamma_{ph}(\Delta)]^2}.
\end{equation}
We also note that for arbitrary relation between $\Gamma_0$ and $\Gamma$ the radiative damping should be included only in the denominator of Eq.~\eqref{Abs:tot:App}, see Eq.~\eqref{abs:dir:1} and Ref.~\cite{Ivchenko:2005a}. At $\Delta \lesssim \Gamma$ the general expression~\eqref{Abs:tot:App} passes to the direct absorbance, Eq.~\eqref{abs:dir:1}. At $\Delta \gg \Gamma$ Eq.~\eqref{Abs:tot:App} provides the indirect transitions contribution derived in Eq.~\eqref{abs:indir:2}. The photon self-energy is given by, up to a factor $\propto |M_{opt}|^2$, by the dressed exciton Greens function. 

It is interesting to note that the approach developed above can be also applied for excitons interacting with free carriers or static disorder, if any. For example, for exciton interacting with short-range impurities the $\gamma_{ph}(\Delta)$ in Eqs.~\eqref{self:1} and \eqref{Abs:tot:App} should be replaced by the Born scattering rate $\gamma_{imp}$ (proportional to the impurity density and the electron-impurity scattering rate) which is independent on the exciton energy. 

\section{Comparison with conventional quasi-two dimensional semiconductors}\label{appendix:comp}

The analysis shows that exciton-phonon interaction is quite strong in transition metal dichalcogenide MLs as compared with conventional GaAs-based quantum wells. For example the parameter $c_1$ in GaAs quantum wells~\cite{PhysRevLett.115.267402} is an order of magnitude smaller than in MoSe$_2$ despite higher values of deformation potential and the fact that exciton interacts with bulk (3D) phonons in GaAs-case. Qualitatively, the enhanced exciton-phonon interaction in atom-thin crystals is due to the fact that for small wavevectors the phonon density of states is larger in 2D than in 3D. 
One can demonstrate this difference between TMD MLs and quantum well structures like GaAs/AlGaAs by determining the phonon scattering rate, as in Eq.~\eqref{lifetime:Temp} but in this case for two-dimensional excitons interacting with bulk (3D) phonons:
\begin{equation}
\label{lifetime:Temp:3D}
\frac{1}{\tau_{T}^{3D}} = \frac{2\pi}{\hbar} \sum_{\bm k'\bm q} |M^{3D;\bm q}_{\bm k'\bm k}|^2 (1+2n_q) \delta (E_k - E_{k'} - \hbar\Omega_q).
\end{equation}
Here
\begin{equation}
\label{M:em:3D}
M^{3D;\bm q}_{\bm k'\bm k} = \sqrt{\frac{\hbar}{2\rho_{bulk} \Omega_{ q}V}} q (D_c - D_v) \mathcal F(q_\parallel) \mathcal F_z(q_z) \delta_{\bm k, \bm k'+\bm q_\parallel},
\end{equation}
where $\rho_{bulk}$ is the bulk mass density (gm$/$cm$^3$), $V$ is the normalization volume, $\bm q =(\bm q_\parallel, q_z)$ and the form-factor $\mathcal F_z(q_z)$ accounts for the exciton size quantization along the quantum well growth axis.In Eq.~\eqref{lifetime:Temp:3D} below the sum over $\bm k'$ is removed due to the momentum conservation law, $\bm k' = \bm k - \bm q$, while the sum over $\bm q$ is transformed to the integral in the three-dimensional case as
\[
\sum_{\bm q} = \frac{V}{(2\pi)^3} \int dq_\parallel q_\parallel \int d\phi_{\bm q} \int dq_z,
\] 
where $\bm q_\parallel$ is the in-plane wavevector of the phonon, $\phi_{\bm q}$ is the in-plane angle of the phonon wavevector, and $q_z$ is its normal component. By contrast, in the case of exciton interacting with two-dimensional phonons, the last integral over $q_z$ is absent. As a result, for phonons with small energies $\hbar\Omega_q \ll k_B T$, one has, instead of Eq.~\eqref{lifetime:Temp} 
\begin{equation}
\label{lifetime:Temp:3D}
\frac{1}{\tau_{T}^{3D}} = \frac{(D_c - D_v)^2 M k_B T}{\hbar^2\rho_{bulk} s^2}\int \mathcal F^2_z(q_z)  \frac{d q_z}{2\pi} \sim \frac{1}{\tau_T} \times \frac{a_0}{a},
\end{equation}
where $a_0$ is the lattice constant and $a$ is the well width (we took into account that $\int \mathcal F^2_z(q_z){d q_z}/({2\pi})  \sim 1/a$ and $\rho_{bulk} \sim \rho/a_0$). The parameter $a_0/a$ is indeed small $\sim 0.1\ldots 0.001$ for typical quantum well structures. Its appearance in Eq.~\eqref{lifetime:Temp:3D} as compared with Eq.~\eqref{lifetime:Temp} follows from the simple argument: for bulk phonons the phonon density of states is $q^2\to  q_z q_{\parallel} \sim q_{\parallel}/a$.

\end{document}